# HSD-PAM: High Speed Super Resolution Deep Penetration Photoacoustic Microscopy Imaging Boosted by Dual Branch Fusion Network

Zhengyuan Zhang, Haoran Jin, Zesheng Zheng, Wenwen Zhang, Wenhao Lu, Feng Qin, Arunima Sharma, Manojit Pramanik, Yuanjin Zheng

*Abstract*—Photoacoustic microscopy (PAM) is a novel implementation of photoacoustic imaging (PAI) for visualizing the 3D bio-structure, which is realized by raster scanning of the tissue. However, as three involved critical imaging parameters, imaging speed, lateral resolution, and penetration depth have mutual effect to one the other. The improvement of one parameter results in the degradation of other two parameters, which constrains the overall performance of the PAM system. Here, we propose to break these limitations by hardware and software co-design. Starting with low lateral resolution, low sampling rate AR-PAM imaging which possesses the deep penetration capability, we aim to enhance the lateral resolution and up sampling the images, so that high speed, super resolution, and deep penetration for the PAM system (HSD-PAM) can be achieved. Data-driven based algorithm is a promising approach to solve this issue, thereby a dedicated novel dual branch fusion network is proposed, which includes a high resolution branch and a high speed branch. Since the availability of switchable AR-OR-PAM imaging system, the corresponding low resolution, undersample AR-PAM and high resolution, full sampled OR-PAM image pairs are utilized for training the network. Extensive simulation and *in vivo* experiments have been conducted to validate the trained model, enhancement results have proved the proposed algorithm achieved the best perceptual and quantitative image quality. As a result, the imaging speed is increased 16× times and the imaging lateral resolution is improved 5× times, while the deep penetration merit of AR-PAM modality is still reserved.

## I. INTRODUCTION

Photoacoutic (PA) Imaging or Optoacoustic (OA) Imaging relies on the physical process of PA effect, which is known for generating imaging results with both optical contrast and acoustic penetration depth [1]. In biomedical PA imaging (PAI) experiments, external pulsed/modulated laser is used to irradiate the bio-tissue, where the energy is absorbed and partially transferred into heat. Thus, the acoustic wave can be generated via thermoelastic expansion and can be detected by ultrasound transducer, so that imaging results can be formed via reconstruction algorithm [2-3]. The amplitude of the acoustic signal is dependent on the local optical energy absorption. The PAI system can be categorized into two major implementations, *i.e.,* PA computed tomography (PACT) and PA microscopy (PAM). The PACT system utilizes multi-element ultrasound transducer arrays to realize real-time (5–20 Hz) and deep penetrate (up to ~5 cm) imaging capability [4], however, it is hard to acquire high resolution imaging in finer structures [5]. In contrast, the PAM system employs single-element ultrasound transducer to receive the acoustic signal in each scanning point. The whole system is driven by two step motors in X and Y axis respectively, so that raster scanning of X-Y plane is enabled. In this manner, a 3D volumetric C-scan imaging results can be acquired, which reveals the anatomy of underlying bio-structure. Conventionally, the PAM system can be further classified into two types, *i.e.*, optical resolution PAM (OR-PAM) and acoustic resolution PAM (AR-PAM) [6].

The OR-PAM is used for superficial tissue (< 1 mm) imaging which confined within the optical focusing limit, thereby the imaging resolution is determined by optical focal spot (< 10 μm) [7]. Beyond the optical focusing limit, the optical beam will be significantly scattered, which results in its loss of effectiveness. What's more, to maintain its high resolution imaging quality, the scanning step size should be kept low (2~4 μm), which also slow down the imaging speed. In conclusion, the high resolution (< 10 μm) imaging of OR-PAM comes at the cost of shallow penetration depth and low imaging speed.

By contrast, the AR-PAM can be used for deep tissue (3-10 mm) imaging which relies on the acoustic focusing, since the

Manuscript received May 16, 2023; This research is supported by the Ministry of Education, Singapore, under its MOE ARF Tier 2 (Award no. MOE2019-T2-2-179). Any opinions, findings and conclusions or recommendations expressed in this material are those of the author(s) and do not reflect the views of the Ministry of Education, Singapore. (Corresponding author: Yuanjin Zheng.).

Zhengyuan Zhang, Zesheng Zheng, Wenwen Zhang, Wenhao Lu, Feng Qin, and Yuanjin Zheng are with the School of Electrical and Electronic Engineering, Nanyang Technological University, Singapore (e-mail: zhengyua002@e.ntu.edu.sg; zesheng001@e.ntu.edu.sg; wenwen.zhang@ntu.edu.sg; wenhao.lu@ntu.edu.sg; n2107549j@e.ntu.edu.sg; yjzheng@ntu.edu.sg).

Haoran Jin is with Zhejiang University, College of Mechanical Engineering, The State Key Laboratory of Fluid Power and Mechatronic Systems, Hangzhou, China (e-mail: jinhr@zju.edu.cn).

Arunima Sharma is with Johns Hopkins University, Electrical and Computer Engineering, Baltimore, USA (e-mail: asharm95@jhu.edu).

Manojit Pramanik is with Iowa State University, Department of Electrical and Computer Engineering, Ames, Iowa, USA (e-mail: mano@iastate.edu).

acoustic diffusion (∼0.0012/mm at 5 MHz) in bio-tissue is much weaker than optical scattering (∼10/mm at 700 nm). However, the acoustic focusing spot is usually ten times wider than optical focusing spot, which result in the inferior spatial resolution (45 μm ∼ 1.16 mm) [7-12]. Nevertheless, the low spatial resolution of AR-PAM imaging also loosens the constraint to the scanning step size (15~100 μm), which speed up the scanning process when the same area is imaged [13-14]. Thereby except for the low imaging lateral resolution, AR-PAM possesses superior penetration depth and less scanning time over the OR-PAM modality.

Recently, deep learning and data driven based methods have been prevailed in industry application [15-16] and academic research, which have also been widely used for PA image reconstruction [17-23] and enhance the raw imaging results [24-27]. Some researchers focus on using the algorithm to improve the imaging speed; For example, DiSpirito, Anthony, *et al.* proposed to use Fully Dense U-Net (FDU-Net) to enhance the downsampled OR-PAM images [28]. Further, Vu, Tri, *et al.* proposed a self-supervised algorithm to up sample the OR-PAM image by exploiting deep image prior [29]. Recently, Seong, Daewoon, *et al.* proposed a modified super-resolution ResNet (SRResNet) to upsample the 3D volumetric PA data, which achieved 80-times faster-imaging speed and 800-times lower data size [30]. While some other researchers aim to use the deep learning techniques to improve the PA imaging quality (*i.e.* lateral resolution). For example, Zhang *et. al* proposed to realistically enhance the lateral resolution of AR-PAM image with MultiResU-Net and conditional generative adversarial network [31-35]. Feng, Fei, *et al.* proposed to use convolutional neural network to implement the deconvolution operation to the AR-PAM images, so that the lateral resolution of the imaging results can be improved [36]. Zhou, Yifeng *et al.* proposed to use conditional GAN to enhance the imaging resolution obtained with Bessel beam to similar level of Gaussian beam, while overcoming the limited depth of focus imposed by the Gaussian-beam excitation [37]. However, none of previous works have explored using algorithm to enhance the imaging speed and imaging resolution simultaneously. In fact, the imaging step size (which determines the imaging speed under the same laser pulse repetition rate) and imaging lateral resolution are two corelated imaging parameters, which have mutual effect to one the other. Decreasing the imaging step size can potentially increase the upper bound of the achievable imaging lateral resolution according to the Nyquist sampling theorem, thus making it possible to realize the super resolution imaging capability. Here a dual branch network is designed which can simultaneously enhance the imaging speed and imaging resolution within an unified framework given low resolution undersampled AR-PAM image as input.

To simultaneously integrate the high resolution capability of OR-PAM imaging with the high speed & deep penetration properties of AR-PAM imaging, we proposed a dual branch fusion network for building a hybrid PAM imaging system based on software and hardware co-design. By equipping the original AR-PAM system with the proposed algorithm, a high speed, super resolution, deep penetration PAM (HSD-PAM) system is built, which possesses advantages over the sole OR-PAM or AR-PAM system. Compared with pure OR-PAM system, the hybrid PAM system has the advantage of much higher imaging speed and deeper penetration depth; Compared with pure AR-PAM system, the hybrid PAM system can achieve significantly higher imaging resolution even breaking the physical limit of lateral resolution that can be achieved by AR-PAM modality. Finally, both simulation and *in vivo* experiments have been conducted to verify the capability of the proposed algorithm as well as the comprehensive performance of the hybrid PAM system. The HSD-PAM system can achieve deep penetration imaging with 5× lateral resolution enhancement and 16× speed up based on experimental results.

## II. DEGRADATION MECHANISM AND RESTORATION ALGORITHM

To develop an algorithm for enhancing low resolution undersampled AR-PAM image to high resolution full sampled OR-PAM image, the degradation mechanism of AR-PAM imaging is first explored. The degradation model can be exploited to generate huge amounts of training image corpus given the ground truth OR-PAM images. Further, a novel neural network specifically for solving the enhancement problem is proposed, which is a dual branch model to enhance the resolution in the high resolution branch and upsample the images in the high speed branch. Based on the generated dataset and developed neural network model, the training strategy for model fitting is also developed and presented.

### A. Degradation Model

Compared to high quality OR-PAM imaging result, the low quality AR-PAM imaging result suffers from two major degradation mechanisms, *i.e.* blurring caused by point spread function (PSF) and down sampling caused by low sampling rate along both X and Y axis.

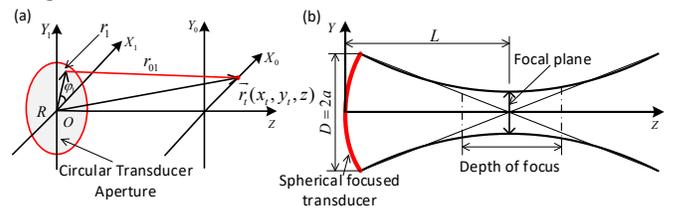

Fig. 1. Schematic of plane circular transducer and spherical focused transducer. (a) Field pattern of plane circular transducer; (b) Beam forming with spherical focused transducer.

The quality of acoustic imaging is controlled under the ultrasound beamforming, by which the lateral resolution is also determined. Thereby the lateral resolution degradation characterized by PSF kernel can also be derived with ultrasound beamforming theory, which can be mathematically expressed with Rayleigh-Sommerfeld equation [38]. Specifically, the field at spatial point $r_t$ generated by the ultrasound transducer can be obtained by adding all the contributions of spherical waves emitted by the points in the transducer surface, which is derived as the following formula:

$$\tilde{E}(\vec{r_t}, \omega) = \frac{1}{i\lambda} \int_0^R \int_{-\pi}^{\pi} \tilde{\epsilon}(r_1, \omega) e^{i\omega r_{01}/c} \frac{z}{r_{01}^2} r_1 d\varphi_1 dr_1$$
$$+ \frac{1}{2\pi} \int_0^R \int_{-\pi}^{\pi} \tilde{\epsilon}(r_1, \omega) e^{i\omega r_{01}/c} \frac{z}{r_{01}^3} r_1 d\varphi_1 dr_1, \quad (1)$$

where the first term characterizes the high frequency portion while the second term characterizes the low frequency portion. Obviously, it is difficult to obtain its analytical solution with its two double-fold integrals. Simplification can be made in practical beamforming scenarios, the second term is negligible so that this term is skipped.

For an unfocused ultrasound transducer, as presented in Fig. 1(a), the produced beam has the beamwidth as large as the diameter of the transducer within the Rayleigh distance, which results in the low resolution of the obtained imaging results. Practically for biomedical PAI experiments, spherical focused transducer, as presented in Fig. 1(b), is used to improve the beamwidth near the imaging focal plane, thus the imaging lateral resolution is also enhanced. Due to its spherical shaped surface, the weighting function (spherical phase shifter) in Eqn. (1) should be further specified as:

$$\tilde{\epsilon}(r_1, \omega) = e^{-i\omega(\sqrt{L^2 + r_1^2} - L)/c}, \quad (2)$$

where $L$ is the focal length. Substituting the Eqn. (2) into Eqn. (1), the field pattern in the focal plane can be analytically derived by calculating double-fold integral as:

$$\tilde{E}(\vec{r_t}, \omega) = \frac{\omega a^2}{icz} e^{i\omega\left(z + \frac{x_t^2}{2z}\right)} \left[\frac{2J_1(\omega x_t a/c)/z}{(\omega x_t a/c)/z}\right]. \quad (3)$$

The PSF kernel can be calculated by taking the integral of the field pattern with frequency spectrum $T(\omega)$ towards the whole bandwidth:

$$h = W(\vec{r_t}) = \int T(\omega) \tilde{E}(\vec{r_t}, \omega) d\omega. \quad (4)$$

Due to the difference of lateral resolution between OR-PAM and AR-APM, the scanning step size and sampling rate are also correspondingly adjusted. The size of AR-PAM imaging results can be considered as downsampled from its OR-PAM counterparts. Assuming that downsampling ratio along X-axis is $d_x$, the downsampling ratio along Y-axis is $d_y$, thus the transformation between results of OR-PAM and AR-PAM can be mathematically expressed as:

$$y = (h \otimes x)_{\downarrow d_x, \downarrow d_y} + n, \quad (5)$$

where $x$ is the high resolution full sampled OR-PAM imaging result, $y$ is the low resolution under sampled AR-PAM imaging result, $n$ is the additive noise associated with AR-PAM imaging, which is approximated as additive white gaussian noise (AWGN) with the standard deviation of $\sigma$.

### B. Network structure

To enhance the low resolution under sampled AR-PAM image $y$ to high resolution full sampled OR-PAM image $x$, the provided training dataset needs to be fully employed to guide the reconstruction process. As has been described earlier, the obtained AR-PAM image suffers from two major degradations, *i.e.* low resolution and low sampling rate. Therefore, a dual branch neural network is proposed and developed, which includes both high resolution branch and high speed branch [39]. The ground truth OR-PAM image is also downsampled into the same size as AR-PAM image to assist another branch of training. Specifically, the downsampled OR-PAM and AR-PAM image pairs are used to supervise the training of high resolution branch; the original high resolution OR-PAM and AR-PAM image pairs are used to supervise the training of high speed branch. The feature maps generated by these two branches are fused together by a gated module, the complementarily fused information is employed for reconstructing the final image. The schematic of proposed network's structure is illustrated in Fig. 2. The detailed description of the architecture and its working principle for each branch/component are explained in the following.

*1)* **High resolution branch**: The high resolution branch in the proposed network is utilized to remove the blurring effect introduced by the acoustic PSF kernel, which is trained by the AR-PAM and downsampled OR-PAM image pairs. Overall, the high resolution branch has a encoder-decoder structure. The encoder part consists of 3 local Resblocks with 2 intermediate convolutional downsampling operators (stride of 2), each local Resblock contains 3 sub-resblocks, which are densely connected to avoid gradient vanishing and fuse the feature maps together for processing. The sub-resblock is made up with two convolutional layers and one ReLU layer. The decoder part contains two deconvolution blocks to upsample the downsampled feature map to recover the feature map to the same as original input images. Each deconvolution block consists of a transposed convolutional layer, a convolutional layer and a ReLU layer. The output feature maps of the deconvolution blocks have two flowing directions. The first direction goes to two convolutional layers with a global residual connection and construct the resolution enhanced output; Another direction is fused with other feature maps in the gated module for further processing.

*2)* **High speed branch**: The high speed branch is supervised by AR-PAM image and the full sampled OR-PAM image, which upscales the feature maps while training. This branch consists of three components: feature extraction component, gated fusion component, and reconstruction upsampling component.

**a) Feature extraction component:** The feature extraction component aims to extract low level features from the input AR-PAM images for reconstructing the high resolution full sample OR-PAM image. This component starts with a convolutional layer followed by a leakyReLU layer, where the leaky slope equals to 0.1. The middle of this branch consists of 4 resblocks, with two convolutional layers and a leakyReLU layer in each resblock, the resblocks are also densely connected. The obtained feature map then goes to a convolutional layer, which will be fused at the gated fusion module.

**b) Gated fusion component:** the gated fusion component is designed to fuse the feature maps from the high resolution branch, feature extraction component with the input AR-PAM image. The feature maps from the high resolution branch have good response on high level structure while losing the finer details; The feature maps from the feature extraction component reserve the low level structural information while lack the clear contour around the blurring region. Therefore, the combination of high resolution branch and feature extraction component produce a complementary feature maps for further reconstruction.

Specifically, the feature maps from high resolution branch $FM_{HR}$, feature extraction component $FM_{EX}$, and input AR-PAM image $I_{AR}$ are concatenated together as an unified feature map, which is further processed by two convolutional layers and a leakyReLU layer, the above process is abstracted as $G()$ operator. The output is a scoremap, which is multiplied element wisely with the feature map from high resolution branch $FM_{HR}$ then plus the feature map from feature extraction component $FM_{EX}$. All the above processing can be clearly formularized as the following mathematical equation:

$$FM_{FU} = G(FM_{HR}, FM_{EX}, I_{AR}) \times FM_{HR} + FM_{EX}. \quad (6)$$

**c) Reconstruction and Upsampling component:** the final reconstruction and upsamling component takes the complementary feature map generated by gated fusion component $FM_{FU}$ and aims to reconstruct the high resolution full sampled output by upscaling the feature map twice. This module contains 4 resblocks and 2 pixel shuffling layers, the resblock consists of 2 convolutional layers and 1 ReLU layer. The pixel shuffling layer has the upscale factor of 2, which reduces the number of channels while increase the spatial resolution. Finally, the feature maps are processed by convolution and leakyReLU combined operator twice to reconstruct the final high resolution full sample output.

Due to its dual branch architecture supervised by both downsampled and full sampled OR-PAM image simultaneously, the training loss comes from two sources. The first loss $Loss_1$ is calculated between output of high resolution branch $I_{HR}$ and downsampled ground truth OR-PAM image $I_{DGT}$ as:

$$Loss_1 = L(I_{HR}, I_{DGT}); \quad (7)$$

The second loss $Loss_2$ is obtained between the output of high speed branch $I_{RE}$ and the ground truth OR-PAM image $I_{GT}$ as:

$$Loss_2 = L(I_{RE}, I_{GT}); \quad (8)$$

Considering the requirement to enhance both pixel-wise fidelity and structural similarity, both mean squared error (MSE) and structural similarity index measure (SSIM) error are employed to measure the deviation. Specifically, the loss function $L$ can be expressed as:

$$\begin{aligned} L(I_1, I_2) &= L_{MSE}(I_1, I_2) + \tau L_{SSIM}(I_1, I_2) \\ &= ||I_1 - I_2||_2^2 + \tau(1 - SSIM(I_1, I_2)), \quad (9) \end{aligned}$$

where $\tau$ is the weighting parameter. The total loss is a combination of both $Loss_1$ and $Loss_2$, which is balanced by a weighting parameter $\lambda$. Therefore, the total loss $Loss$ is represented as:

$$\begin{aligned} Loss &= Loss_1 + \lambda Loss_2 \\ &= L(I_{HR}, I_{DGT}) + \lambda L(I_{RE}, I_{GT}). \quad (10) \end{aligned}$$

### C. Training details

The capability of the proposed network is gradually enforced by training the model in different steps with simulated and *in vivo* dataset. Since the scarcity of well labelled training data obtained with real imaging system, we first generated large amount of simulated data based on the derived degradation model.

The high resolution, fully sampled ground truth OR-PAM data are acquired from the Photoacoustic Imaging Lab in Duke University [28]. The provided images are first cropped into uniform size of 256×256. The corresponding low resolution, down sampled AR-PAM images are synthesized with Eqn. (5), which has the size of 64×64. In this process, totally 5000 paired images are acquired for training.

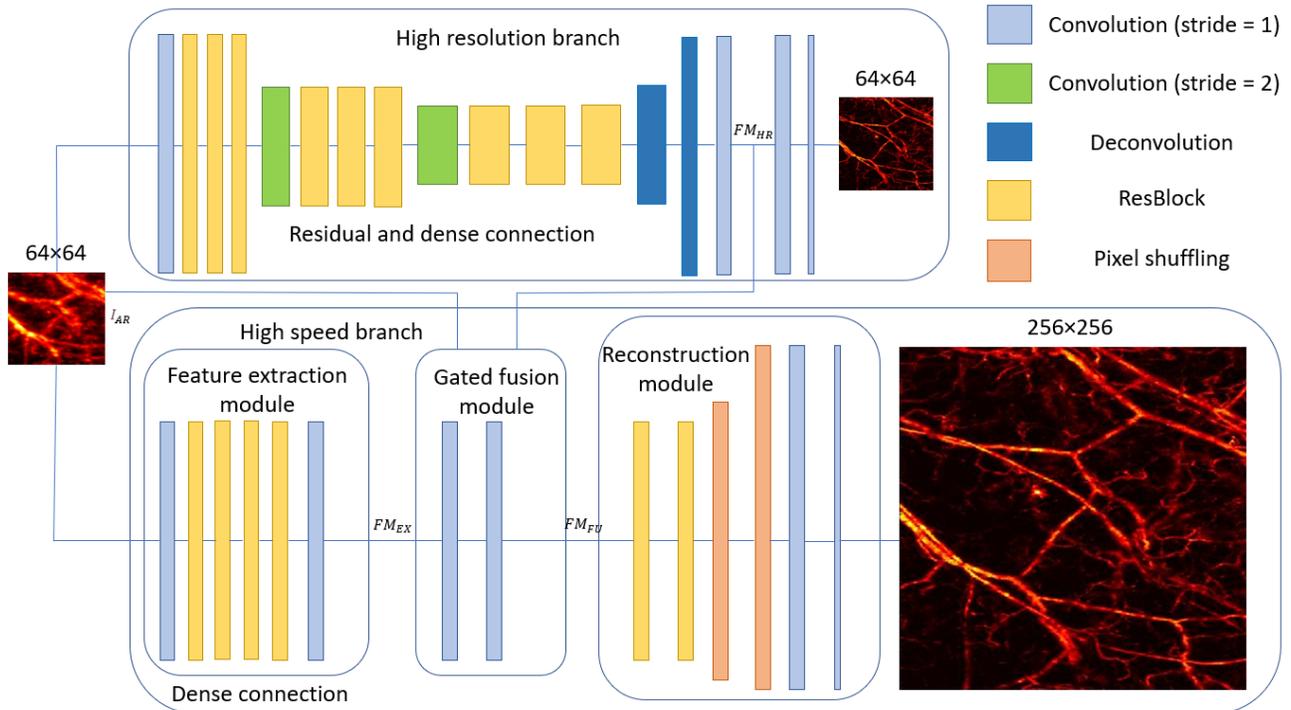

Fig. 2. Schematic of dual branch fusion network architecture and its components for HSD-PAM imaging.

After trained with simulated dataset, the model already achieved high performance to enhance the simulated AR-PAM images to high resolution full sampled OR-PAM images. However, when applied to AR-PAM images obtained from real imaging system, the performance is still unsatisfactory due to the existed domain gap between simulation and *in vivo* situation. To further promote the model's ability to enhance *in vivo* AR-PAM images, we have specifically conducted *in vivo* experiments to generate the real data for training. Actually, the *in vivo* data fine-tuned the parameters of the previous pre-trained model and further improve the performance from the baseline trained with simulated dataset. After fine tuning with *in vivo* data, the model is further optimized to fit for enhancing *in vivo* AR-PAM images with limited real imaging data, which also justified the necessity of this two-step training approach.

In terms of training hyperparameters, the training epoch is selected as 50, and the learning rate is altered follow a pre-defined function, which is selected as $10^{-4}$ in the first 25 epochs and decreased to $10^{-5}$ after 25 epochs. The batch size is determined as 4, which will be shuffled in every new training epoch. The ADAM algorithm is selected as the optimizer to update the neural network's parameters, and the coefficients used for computing the average gradient and squared gradient are selected as $\beta_1 = 0.9$ and $\beta_2 = 0.999$. All the code is developed under the PyTorch library in python, the training process is accelerated with NVIDIA 1080Ti GPU.

## III. Switchable AR-OR-PAM Imaging System

To fine tune the model and generate data for validation, *in vivo* data acquired from real imaging system are required. In this work, the *in vivo* experiments were conducted by a switchable AR-OR-PAM imaging system, the detailed information and schematic of the system can be found in [14]. The excitation source is a nanosecond tunable laser system, which includes a diode-pumped Nd:YAG laser (INNOSLAB, Edgewave, Wurselen, Germany) followed by a dye laser (Credo-DYE-N, Sirah dye laser, Spectra Physics, Santa Clara, CA, USA), with which the wavelength of the laser can be adjusted. In this work, all the experiments were operated at the laser wavelength of 570 nm with a maximum repetition rate of 5000 Hz. A small portion (5%) of laser beam was diverted by beam splitter to the photodiode and recorded with first channel of oscilloscope, which is also utilized as control signal for synchronization of data acquisition and stage motion. Using a right-angle prism placed on a motorized stage (CR1/M-Z7, Thorlabs), the beam can either be diverted towards a multi-mode fiber (M29L01, Thorlabs) which is connected to the AR-PAM set-up or can be focused using a fiber coupler (F-91-C1, Newport) on a single-mode fiber (P1-460B-FC-1, Thorlabs) which is connected to the OR-PAM set-up. For AR-PAM, the fiber output is passed through a conical lens and then weakly focused on the sample using a homemade optical condenser (cone angles: $70^o, 110^o$). The fiber output from single mode fiber is collimated and further focused strongly on the target tissue using achromatic lenses. Confocally aligned focused ultrasound transducers (V214-BB-RM, Olympus-NDT, Waltham, MA, USA) with central frequency of 50 MHz are used to acquire PA signals in both AR- and OR-PAM modes. The ultrasonic transducer was attached with an acoustic lens (LC4573, Thorlabs) with radius of curvature 4.6 mm, diameter 6 mm, and focal length 10 mm to provide an acoustic focal spot of ~46 μm. For acoustic coupling, the bottom of the transducer was submerged in water in a water tank which contained a 10 cm hole covered with thin transparent polyethylene sheet. The signal is amplified by two amplifiers (ZFL-500LN, Mini Circuits, Brooklyn, NY, USA) and collected at 250 Ms/s sampling rate using data acquisition card (M4i.4420, Spectrum, Grosshansdorf, Germany). The AR-OR combined system is connected to a 3-axis motorized stage (PLS 85 for X and Y axis, VT 80 for Z axis, PI—Physik Instrumente, Karlsruhe, Germany), which allows the scanning head to implement continuous two-dimensional raster scanning over the sample and acquire images. The motorized stages were controlled by a 3-axis controller connected to the computer. By sliding the scanning head, it is possible to image the same area of a sample using both the set-ups without physically moving the sample. While conducting the AR-PAM imaging experiments, the scanning step size is determined as 16 μm along both of the X and Y axis. By contrast, the scanning step size for conducting OR-PAM imaging experiments is 4 μm along the X and Y axis. The sampling rate of OR-PAM imaging is four times higher than the AR-PAM imaging, which coincides with the overall upscale factor of the network that we have developed.

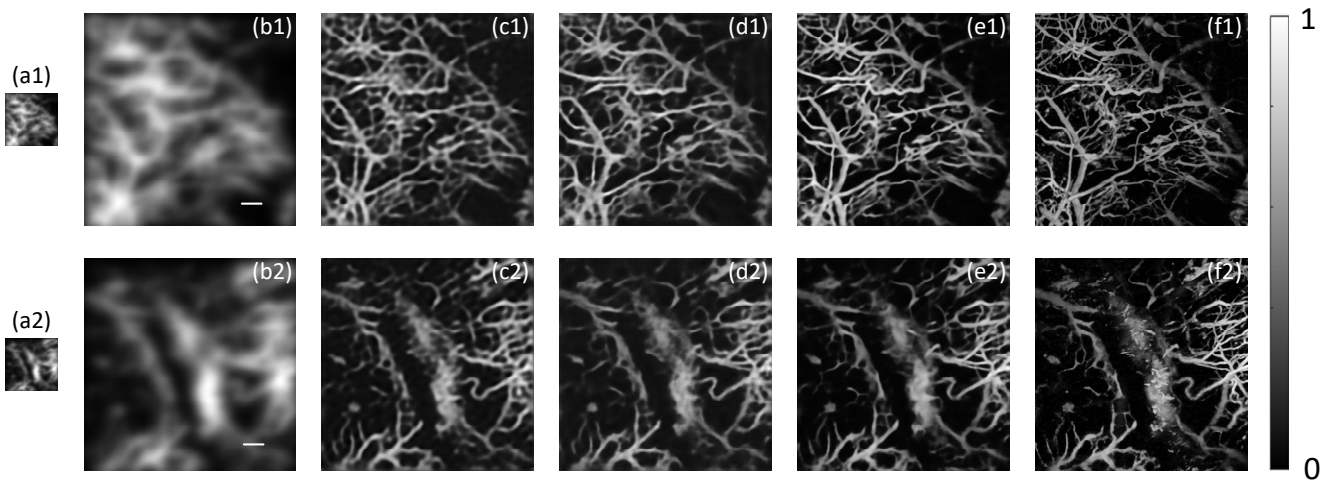

Fig. 3. Enhancement results of two simulated mouse brain AR-PAM images. (a) Low resolution undersampled AR-PAM image; (b) Enhancement results from bicubic interpolation; (c) Enhancement results from FDU-Net; (d) Enhancement results from MultiResU-Net; (e) Enhancement result from proposed algorithm; (f) Ground truth OR-PAM image. Scale bar: 1 mm.

For *in vivo* ear imaging, a mouse (procured from InVivos Pte. Ltd., Singapore) was anesthetized and placed such that its ear was in contact with the polyethylene sheet. Animal experiments were performed according to the approved guidelines and regulations by the institutional Animal Care and Use committee of Nanyang Technological University, Singapore (Animal Protocol Number ARF-SBS/NIE-A0263).

## IV. RESULTS

The effectiveness of the proposed algorithm for achieving high speed, super resolution, deep penetration PAM imaging must be validated and supported by experimental enhancement results. With the derived imaging degradation model and available integrated AR-OR-PAM imaging system, the validation can be implemented in a comprehensive and objective way. In this paper, the results are obtained from both simulation and *in vivo* experiments. In each experiment, the enhancement results generated by bicubic interpolation, FDU-Net, MultiResU-Net, and proposed network are presented. Since the availability of ground truth data in both simulation and *in vivo* experiments, the peak signal-to-noise ratio (PSNR) and SSIM metrics can be directly employed for judging the enhancement performance of the algorithms, which are widely recognized criteria and can reflect the performance in comprehensive perspectives. Besides, since the lateral resolution is a critical parameter in PAM imaging, a more accurate criterion full width at half maximum (FWHM) is utilized here to measure the lateral resolution of the enhanced results [40]. The detailed qualitative analysis and quantitative measurement of the enhancement results are presented in the following.

### A. Enhancement results of simulation data

With the derived degradation model and ground truth OR-PAM imaging data, the proposed model is first trained with synthesized dataset. The trained model is utilized here for simulated AR-PAM image enhancement, two exemplar enhancement results are demonstrated in Fig. 3.

Specifically, the simulated low resolution undersampled AR-PAM image is presented in Fig. 3(a1) & (a2). The enhancement results generated by bicubic interpolation, FDU-Net, MultiResU-Net, and proposed network are presented in Fig. 3 (b1) & (b2), 3(c1) & (c2), 3(d1) & (d2) and 3(e1) & (e2), respectively. The ground truth OR-PAM image for comparison is depicted in Fig. 3(f1) & (f2). As can be seen from the results, the simulated AR-PAM image is under sampled (with small size) and presented low lateral resolution. While the ground truth OR-PAM image is full sampled with large size and high lateral resolution presented abundant finer vasculature details, which also clearly indicates the necessity to enhancement and confirms our research motivation. By applying bicubic interpolation, only the image size is upscaled while the lateral resolution is too low to distinguish any meaningful details. The enhancement results from FDU-Net and MultiResU-Net are upscaled and the lateral resolution have also been significantly improved, however, some finer details are lost and piecewise artifacts are still presented. By contrast, the enhancement result from our proposed algorithm presents the best perceptual quality, all the microvasculature details can be clearly observed. The fidelity of the enhancement can also be identified when compared with the ground truth OR-PAM image.

Quantitatively, the algorithms' performance for simulated

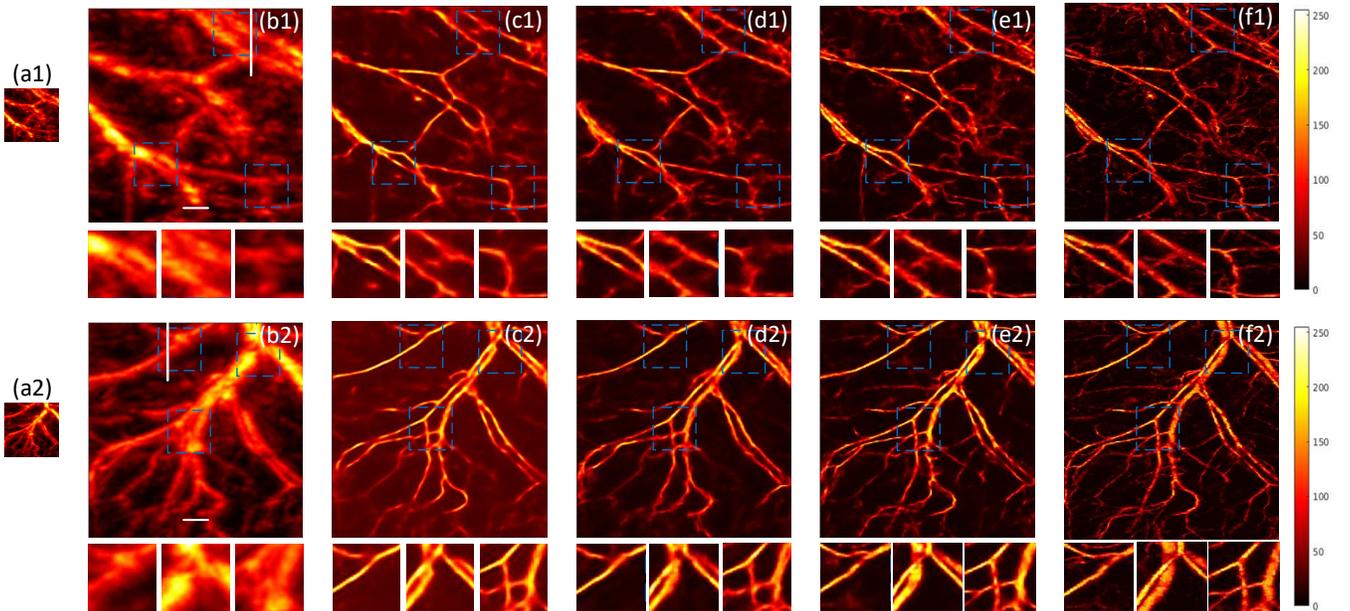

Fig. 4. The enhancement of two *in vivo* AR-PAM imaging results. (a) Low resolution undersampled AR-PAM image; (b) Enhancement results from bicubic interpolation; (c) Enhancement results from FDU-Net; (d) Enhancement results from MultiResU-Net; (e) Enhancement result from proposed algorithm; (f) Ground truth OR-PAM image; Scale bar: 1 mm.

AR-PAM image enhancement is measured with PSNR and SSIM criteria. All the quantitative results are calculated based statistical analysis, the mean value and standard deviation are demonstrated in Table I. As can be seen from the quantitative values, our proposed algorithm consistently achieved the best enhancement results in terms of both lateral resolution and overall imaging quality, which presented highest PSNR/SSIM values and most close lateral resolution to the ground truth OR-PAM image.

TABLE I
STATISTICAL MEASUREMENT OF IMAGE ENHANCEMENT QUALITY.

|  | Bicubic Interpolation | FDU-Net | MultiRes-Net | HSD-PAM |
|---|---|---|---|---|
| PSNR dB (mean±sd) | 13.85 ± 2.17 | 20.93 ± 2.02 | 23.53 ± 1.94 | **24.79 ± 2.20** |
| SSIM (mean±sd) | 0.31± 0.04 | 0.63± 0.06 | 0.70 ± 0.04 | **0.75 ± 0.05** |

### B. Enhancement results of in vivo data

More critically, the proposed algorithm should be applied to enhance the *in vivo* PAM images obtained from real imaging system. The model fine tuned with *in vivo* dataset is employed here for validation. Thanks to the availability of our switchable AR-OR-PAM imaging system, AR-PAM image and its corresponding ground truth image with good alignment can be obtained, thus a fair and objective comparison between the enhancement results is feasible. The input AR-PAM image is presented in Fig. 4(a1) & (a2), while the ground truth OR-PAM image is depicted in Fig. 4(f1) & (f2). for comparison. The enhancement results of input AR-PAM image by Bicubic interpolation, FDU-Net, MultiResU-Net, and the proposed algorithm are presented in Fig. 4(b1) & (b2), 4(c1) & (c2), 4(d1) & (d2) and 4(e1) & (e2). To better verify the performance, three regions of interest (ROI) are marked with blue bounding boxes and used for comparison, which indicate different vasculature structures. Specifically, in ROI 1 vessel bifurcations are presented; in ROI 2 large arteries and the small veins are illustrated; in ROI 3 microvasculature and capillary are depicted. The zoomed in and enlarged area of the ROIs are also presented in Fig. 4. Qualitatively, the enhancement results from proposed algorithm achieved the best perceptual performance that is closest to ground truth OR-PAM image. Both arteries and microvasculature are restored with clear edge and boundaries. The bifurcation of the vessels is also clearly split apart and distinguishable.

TABLE II
QUANTITATIVE MEASUREMENT OF THE ENHANCED *IN VIVO* IMAGE.

|  | Bicubic Interpolation | FDU-Net | MultiResU-Net | HSD-PAM |
|---|---|---|---|---|
| PSNR | 14.16 | 21.93 | 23.61 | **25.62** |
| SSIM | 0.39 | 0.61 | 0.68 | **0.73** |

Since the ground truth OR-PAM image is available, PSNR and SSIM criteria can also be calculated to evaluate the enhancement performance. The obtained quantitative values are summarized and demonstrated in the following Table II. Further, the signal intensity along the vertical line of all the enhancement results are presented in Fig. 5., where we can see the good consistency between the ground truth data and output of the proposed algorithm. Based on the intensity profile, the FWHM value can also be measured, which indicates the lateral resolution of each enhancement result. The calculated lateral resolution in the two cases are summarized and demonstrated in the following Table III (the unit of the measurement is µm). The results proved that the lateral resolution have been enhanced around 5 times by applying the proposed computational techniques.

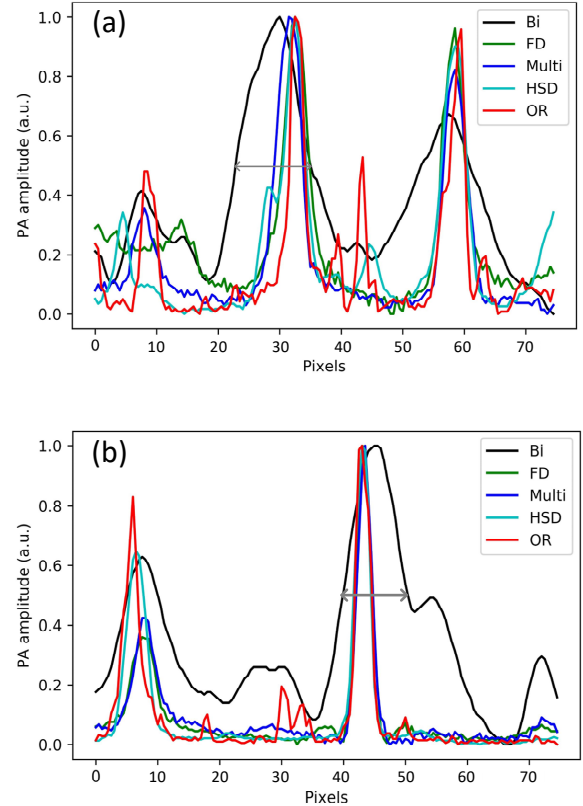

Fig. 5. (a) Signal intensity profiles along the vertical line of position 1; (b) Signal intensity profiles along the vertical line of position 2.

TABLE III
FULL WIDTH AT HALF MAXIMUM VALUES MEASURED IN THREE CASES.

|  | AR-PAM | Bicubic Interpolation | FDU-Net | MultiResU-Net | HSD-PAM | OR-PAM |
|---|---|---|---|---|---|---|
| 1 | 58 | 58 | 20 | 22 | 18 | 14 |
| 2 | 50 | 53 | 11 | 11 | 10 | 10 |

### C. Enhancement results of deep penetrated in vivo data

To test the proposed algorithm's capability in enhancing deep penetration imaging, the proposed algorithm is further employed to restore the *in vivo* imaging result acquired from high penetration depth. The A-scan, B-scan, and C-scan of the imaging result are illustrated in Fig. 6(a), Fig. 6(b), and Fig. 6(c). As can be seen from the B-scan result, the imaging depth spans the depth between 2~6 mm. Therefore, the imaging results contains the information from deep 3D structures. Further, the algorithm is applied to enhance the corresponding 2D MAP image, where the obtained enhancement result is presented in the Fig. 6(d).

By comparing the original imaging result with the enhanced result, we can find the vasculature both in shallow and deep tissue can be enhanced. The resolution has been significantly enhanced, meanwhile the more vasculature details are restored. What's more, the image size is also upscaled 4 times along both X-axis and Y-axis, thereby the imaging time is significantly reduced. Quantitatively, contrast to noise ratio (CNR) and FWHM metrics are used for measuring the enhancement performance of the algorithm. The signal intensity profile along the vertical dashed line is presented in Fig. 6(e), noted that the original image in Fig. 6(c) is upscaled to the same size as Fig. 6(d) for a fair comparison. The quantitative values have clearly reflected the improvement of imaging quality by our algorithm, as depicted in Table IV.

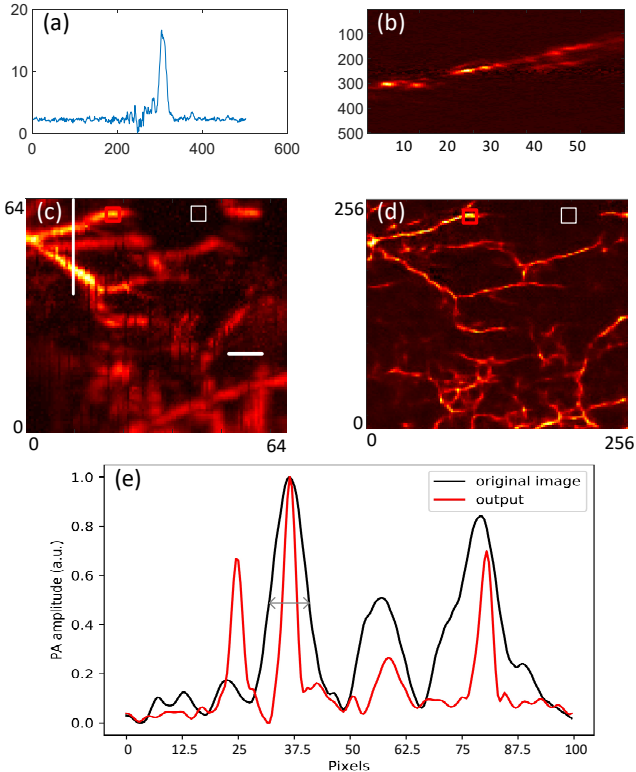

Fig. 6. (a) A-scan of the deep penetrated *in vivo* imaging result; (b) B-scan of the deep penetrated *in vivo* imaging result; (c) 2D MAP image of the deep penetrated *in vivo* imaging result; (d) Enhancement result of the 2D MAP image; (e) Signal intensity profiles along the vertical line. Scale bar: 1 mm.

TABLE IV
QUANTITATIVE RESULTS OF CNR/FWHM CRITERIA FOR IMAGES.

|  | 2D MAP image | Enhanced result |
| --- | --- | --- |
| CNR | 9.38 | 11.57 |
| FWHM (μm) | 49 | 18 |

## V. DISCUSSION

The three critical PA imaging parameters: imaging speed, imaging lateral resolution, and imaging penetration depth are indeed three correlated parameters. Conventionally, the enhancement of one parameter will directly degrade the performance of the other two parameters. Specifically, the imaging scanning rate $s$ can be characterized with the following formula:

$$s = \frac{1}{scanning\ time} = \frac{f_p}{N_x N_y}, \quad (11)$$

where $f_p$ is the pulse repetition rate, $N_x, N_y$ are the number of scanning pixels along X, Y directions, which are inversely proportional to the scanning step size. The maximum allowable pulse repetition rate, *i.e.* critical pulse repetition rate (PRR) $f_p^c$ can be determined as following formula:

$$f_p^c = \frac{v_{sos}}{d}, \quad (12)$$

where $v_{sos}$ is the speed of sound in the bio-tissue, $d$ is the imaging penetration depth. Therefore, with the increase of imaging depth, the scanning rate will be decrease which slows down the imaging speed. On the other hand, when the scanning step size is reduced and the number of scanning pixels is increased, the imaging speed will also be slow down. Furthermore, the lateral resolution is also closely related to the scanning step size. To obtain high lateral resolution imaging result, the scanning step size is supposed to be smaller than half of the lateral resolution as indicated in Nyquist sampling theorem, so that the actual physical resolution can be achieved. Besides, with deeper penetration, the light is scattered which cannot be focused, the lateral resolution is also decreased since it relies on the acoustic focusing rather than optical focusing. Therefore, the corelated relationship between the three PA imaging parameters can be illustrated in Fig. 7. In a word, to acquire large scale deep penetrated 3D volumetric PA imaging data, the sacrifice of imaging speed/lateral resolution is necessary.

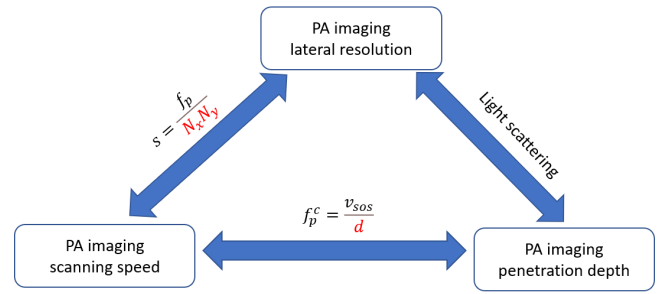

Fig. 7. Relationship between the three PA imaging parameters: lateral resolution, scanning speed, and penetration depth.

To break these physical limitations, we proposed to use computational algorithms to enhance the imaging speed and lateral resolution simultaneously, which also maintains the deep penetration property, so that high speed, super resolution, and deep penetration capability of the PAM system can be achieved simultaneously. Specifically, by introducing a dual branches fusion network, the input under sampled, low resolution AR-PAM image is enhanced by the high resolution branch and high speed branch, the feature maps are further fused together by a gated module to reconstruct the final image. As can be seen from the reconstruction results in both simulation data and *in vivo* data, our proposed algorithm achieved the best

xx

performance qualitatively and quantitatively. Noted that for all the learning based algorithms used for enhancement, the hyper-parameters, i.e. learning rate, training epochs, and batch size are all set as identical while training to guarantee a fair comparison. What's more, since the ground truth data for both simulation and *in vivo* experiments are available, the fidelity of the enhancement results can also be identified.

Based on the enhancement result, detailed performance improvement of the PA imaging parameters can be analyzed. Since the input image is 4× upscaled along both X and Y axis, thereby the imaging speed is 16× improved. Based on the measured FWHM value in the *in vivo* experiment, the lateral resolution is enhanced from 50 μm in the original input AR-PAM image to 10 μm in the reconstructed image, which presented 5 times of enhancement, while the lateral resolution for the ground truth OR-PAM image is also 10 μm. By contrast, when the image is only processed by the high resolution branch but doesn't upscaled, the lateral resolution can only achieve 16 μm, which is still inferior than the ground truth standard. Therefore, we call it super resolution imaging which exceed the physical constraint posed by AR-PAM imaging setup. In terms of penetration depth, the deep penetration capability of AR-PAM modality is still maintained without the sacrifice of imaging speed and imaging lateral resolution. With the above mentioned software and hardware co-design, a high speed, super resolution, and deep penetration PAM (HSD-PAM) system is built.

Nevertheless, it is no doubt that there still exists the space for further exploration and improvement. Firstly, it is not sure what is the up sampling limit of the approach. The scanning time can be further reduced if we further increase the scanning step size. However, what is the lower limit that the proposed algorithm is still capable to restore is unknown. This issue needs to be further investigated in the following work. Secondly, it is still hard to firmly verify the fidelity of the enhancement result acquired from deep tissue due to the unavailability of ground truth data for an objective comparison. But evidence can still be collected from the enhancement results from shallow tissue and the correspondence between the input image and enhanced image, which can assist us to make the judgement that the reconstruction fidelity is guaranteed even in deep tissue.

## VI. Conclusion

This work innovatively proposed to simultaneously enhance the PAM imaging speed and imaging lateral resolution in an integrated framework without trading off deep penetration capability, which significantly improve the imaging performance based on software and hardware co-design. Specifically, a dual-branch fusion network, which includes both high-resolution and high-speed branches, is used to train the paired under sampled low resolution AR-PAM image and full sampled high resolution OR-PAM image, which is a novel approach to solve the issue of improving imaging performance while maintaining deep penetration capability. The extensive conducted simulation and *in vivo* experiments have demonstrated the effectiveness of the proposed method and suggest that this approach has the potential to significantly improve the performance of PAM imaging systems. The imaging speed has improved for 16 times, the imaging lateral resolution has also enhanced around 5 times, while deep penetration merit of AR-PAM imaging is still reserved. Overall, this work represents an important step towards improving the capabilities of PAM imaging and could have broad implications for biomedical imaging research.